\documentclass[12pt,preprint]{aastex}

\def\th{\thinspace}
\def\kms{\ifmmode{\rm km\th s^{-1}}\else km\th s$^{-1}$\fi}

\def\today{\number\year\space \ifcase\month\or  January\or February\or
        March\or April\or May\or June\or July\or August\or
September\or
        October\or November\or December\fi\space \number\day}

\begin{document}

\title{EPSILON AURIGAE: AN IMPROVED SPECTROSCOPIC ORBITAL SOLUTION}

\email{************  To appear in the Astronomical Journal   ***********}

\author{
Robert P. Stefanik\altaffilmark{1},
Guillermo Torres\altaffilmark{1},
Justin Lovegrove\altaffilmark{1,2},
Vivian E. Pera\altaffilmark{3},
David W. Latham\altaffilmark{1}, 
Joseph Zajac\altaffilmark{1}, 
Tsevi Mazeh\altaffilmark{4,5}
}


\altaffiltext{1}{Harvard-Smithsonian Center for Astrophysics, 60 Garden Street, Cambridge, MA 02138}
\altaffiltext{2}{Applied Mathematics Group, University of Southampton, University Road, Southampton, 
SO171BJ, UK}
\altaffiltext{3}{MIT Lincoln Laboratory, 244 Wood Street, Lexington, MA 02420}
\altaffiltext{4}{Wise Observatory, Tel Aviv University, Tel Aviv, 69978, Israel}
\altaffiltext{5}{Radcliffe Institute for Advanced Studies, Harvard University, Cambridge, MA 02138}
\email{rstefanik@cfa.harvard.edu}

\begin{abstract}

A rare eclipse of the mysterious object $\epsilon$ Aurigae will occur
in 2009--2011. We report an updated single-lined spectroscopic
solution for the orbit of the primary star based on 20 years of
monitoring at the CfA, combined with historical velocity observations
dating back to 1897. There are 518 new CfA observations obtained
between 1989 and 2009.  Two solutions are presented.  One uses the
velocities outside the eclipse phases together with mid-times of
previous eclipses, from photometry dating back to 1842, which provide
the strongest constraint on the ephemeris. This yields a period of
$9896.0 \pm 1.6$ days ($27.0938 \pm 0.0044$ years) with a velocity
semi-amplitude of $13.84 \pm 0.23$ km~s$^{-1}$ and an eccentricity of
$0.227 \pm 0.011$. The middle of the current on-going eclipse 
predicted by this
combined fit is JD $2,\!455,\!413.8 \pm 4.8$, corresponding to 2010
August 5. If we use only the radial velocities, we find that the
predicted middle of the current eclipse is nine months earlier. This
would imply that the gravitating companion is not the same as the
eclipsing object. Alternatively, the purely spectroscopic solution may
be biased by perturbations in the velocities due to the short-period
oscillations of the supergiant.

\end{abstract}


\section{INTRODUCTION}

Epsilon Aurigae has been known as a variable system since early in the $19^{th}$ century and as 
an eclipsing system since the beginning of the $20^{th}$ century.  It has the longest known 
period of any eclipsing binary, 27 years.  SIMBAD lists over 400 papers on the system, 
yet the evolutionary state of the system and the properties of neither the primary 
nor the secondary are clearly understood. Even the star's spectral classification, generally
assumed to be F0\,I, is uncertain.
 
The spectroscopic orbit implies that the companion is nearly the same mass as the F 
supergiant primary.  If the primary is a typical F supergiant with a mass of 16~M$_{\sun}$
then the companion has a mass of at least 13~M$_{\sun}$.  However, we, like others before us,
see no light from the 
secondary in our spectra.  It has also been suggested that the primary is an old post-AGB 
star with a mass of less than 3~M$_{\sun}$, implying a secondary mass of 
$\sim$ 6~M$_{\sun}$ \citep{S87}.  

The eclipse 
duration of nearly two years implies a size of the occulting object of many AU.  It 
is generally agreed that the occulting object consists of a complex, donut-shaped, rotating disk 
and an embedded unseen 
object \citep[for a review of the properties of the $\epsilon$ Aurigae system see][]{GD02}.
Various suggestions have been made on the nature of the secondary body, from a black hole to a
protoplanetary system to a pair of stars in a binary system.

The 2009--2011 eclipse has started and a major campaign is underway to observe the system.
For more information on the campaign see
http://mysite.du.edu/ {\Large$\tilde{ }$} rstencel/epsaur.htm 
and http://www.hposoft.com/Campaign09.html \citep[See also][]{HS08}.

To help in the interpretation of the campaign results we re-examine the 
historical radial-velocity data and present the 
results of new radial-velocity monitoring of $\epsilon$ Aurigae in order to determine 
an updated spectroscopic orbit. This provides an improved estimate for the time of 
mid-eclipse of the primary by the gravitating companion responsible for the orbital motion.

\section{RADIAL-VELOCITY OBSERVATIONS}

In order to refine the spectroscopic orbit of the system we have been monitoring 
the radial velocity of $\epsilon$ Aurigae for nearly 20 years at the Harvard-Smithsonian 
Center for Astrophysics (CfA) starting
in November 1989.  We have used the CfA Digital Speedometers \citep{L85,L92} on 
the 1.5-m Wyeth Reflector at the Oak Ridge Observatory (now closed) in Harvard,
Massachusetts and the 1.5-m Tillinghast Reflector at the F.\ L.\ Whipple Observatory 
on Mt.\ Hopkins, Arizona..  These instruments used intensified 
photon-counting Reticon detectors on nearly identical echelle spectrographs, recording 
a single echelle order giving a spectral coverage of 45~\AA\ centered at 
5187~\AA.  Radial velocities were derived from the observed spectra using the
one-dimensional correlation package {\bf r2rvsao} \citep{Kurtz1998} running
inside the IRAF\footnotemark[1] \footnotetext[1]{IRAF (Image Reduction and
Analysis Facility) is distributed by the National Optical Astronomy
Observatories, which are operated by the Association of Universities for
Research in Astronomy, Inc., under contract with the National Science
Foundation.}environment.  For most stars observed with the CfA Digital
Speedometers we use templates drawn from a library of synthetic spectra
calculated by Jon Morse using Kurucz models \citep{Latham2002}.  For each star
we first run grids of correlations over an appropriate range of values in
effective temperature, surface gravity, metallicity, and rotational velocity,
and then choose the single template that gives the highest value for the peak of
the correlation coefficient, averaged over all the observed spectra.  In the
case of $\epsilon$ Aurigae, when we analyzed the 489 CfA spectra available in
September 2008, we found that the best correlation was obtained for effective
temperature = 7750 K, log surface gravity = 1.5 (cgs) and line broadening =
41 \kms, assuming solar metallicity.  However, this template spectrum is not a
particularly good match to the observed spectra, as indicated by the fact that
the average peak correlation value was only 0.80 and the average internal error
estimate was 3.0 \kms. We experimented with the use of observed spectra as the
template, and found that a spectrum obtained on JD 2447995 gave a median
internal error estimate of 0.75 \kms.  Therefore we used that observation as the
template for all the velocity determinations, and adjusted the velocity zero
point so that we got the same radial velocity as yielded by the synthetic
template for that particular observation.  Thus the velocities from the CfA
Digital Speedometers which we report for $\epsilon$ Aurigae in this paper should
be on (or at least close to) the native CfA system described by
\citet{Stefanik1999}.

The CfA radial velocities as well as the other 
historical velocities (see below) are given in Table~\ref{tab:velocities}. In 
this table we report, as examples, the individual heliocentric radial 
velocities for a few selected dates; the entire velocity table is available 
in the electronic version of the Journal.

Epsilon Aurigae's radial velocity has been measured several hundred times by
others over the years and, we 
have gathered the previously reported radial velocities in 
Table~\ref{tab:velocities} along with the CfA velocities.  The major sources of the 
reported radial velocities are:
Potsdam, \citet{L24}: 186 observations covering 4176 days, these included four velocities from
other observatories that were not reported in other publications; 
Yerkes, \citet{FSE29}: 367, 11856 days, these included 14 velocities in a footnote; and
Mt. Wilson, \citet{SPZ58}: 123, 10603 days, these included 14 velocities from DAO Victoria.
Our listed velocities and errors for Mt. Wilson are the mean velocities of all the 
spectral lines reported in the paper 
\citep{SPZ58} and their standard deviation. A few other velocities reported in 
Table~\ref{tab:velocities} are from \citet{CM28}, \citet{A70}, 
\citet{P83}, \citet{BE86}, \citet{C91}, \citet{LS86}, \citet{S87}, \citet{AF85}
 and \citet{BLM86}.
We note that many of these velocities are from individual spectral lines obtained during eclipse
phases and are attributed to the gas disk.

Combining the CfA and previously published velocities yields a total of 1320 
velocities covering over 112 years.  In Figure 1 we 
show all the velocities from Table~\ref{tab:velocities}. Also shown are the previous eclipse intervals 
along with the predicted 2009--11 
eclipse interval.  In Figure 1 
we see the long period orbital radial velocity variation; the short term, lower 
amplitude, irregular atmospheric oscillation of the primary, and the complex velocity 
variation during eclipse phases.  Also shown in Figure 1 is the velocity curve 
calculated using our combined orbital 
solution (see below).

Combining these data requires consideration of zero-point corrections between the
different data sets. However, this
is not straightforward because of the atmospheric oscillations and 
because the various measurements were done at different observatories, 
by different observers, with different instruments 
and different techniques and data reduction.  Furthermore, the radial velocity determinations 
for this star have large errors because $\epsilon$ Aurigae is a hot, rapidly rotating star 
with a line broadening of 40 \kms.
Because of these difficulties no zero-point offsets have been applied when combining the 
different data sets.

\section{KEPLERIAN SPECTROSCOPIC ORBIT}

The first orbital solution was by \citet{L24}. \citet{KSS37} 
combined Yerkes velocities with Ludendorff's Potsdam velocities to update the 
orbital solution. \citet{M62} added observations from Mt. Wilson and DAO 
and calculated a new orbital solution. 
\citet{W70} recalculated the solution using the data reported by Morris.  
In Table~\ref{tab:elements} we give the previously computed orbital solutions, 
where we give 
the period $P$ in days,
the eccentricity $e$, 
the longitude of periastron $\omega$ in degrees, 
the heliocentric Julian date of periastron passage $T - 2,\!400,\!000$, 
the center-of-mass velocity $\gamma$ in \kms, 
the observed orbital semi-amplitude $K$ in \kms, 
the projected semi-major axis a$_1$$\sin{\it{i}}$ in $10^{6}$ km, 
the mass function \it{f}(\it{m}) \rm{in} \rm{M}$_{\sun}$,
the rms velocity residuals $\sigma$,  
the number of observations $N$, and 
the time span both in days and in the number of periods covered.

We note that we have not been able to reproduce the values of $\omega$
reported by these authors.  Re-computing the orbital solutions using 
only the Potsdam, Yerkes and Mt. Wilson velocities, and various combinations of 
these data sets, give $\omega$ values more consistent with the results 
we report below.

In computing an updated spectroscopic orbit we have excluded velocities taken 
during the eclipse phases and 200 days before and after first and last 
contacts.  During these time intervals the structure of the spectral lines
is rather complex, often asymmetric, often multiple, and varies with time with
contributions from both the primary and the secondary disk 
\citep[see e.g.][]{SPZ58,LS86,S87}. The excluded 
Julian Date intervals are: 1901--03 eclipse, JD 2415294--16362; 1928--30 eclipse, 
JD 2425193--26261; 1955--57 eclipse, JD 2435091--36158; 
1982--84 eclipse, JD 2444973--46041 and 2009--11 eclipse, JD 2454880--55948.  
We have also excluded the out-of-eclipse H$\alpha$ velocities because
of their complex structure \citep{C91,Sch07}.  The excluded velocities are shown 
in Figure 1 as open circles.

This yields a new updated Keplerian spectroscopic orbital solution based on 1014
velocities and with the orbital elements shown in the next-to-last column of 
Table~\ref{tab:elements}. The 
rms velocity residual from the orbital fit is 4.6 \kms, dominated by the 
radial oscillations of the atmosphere.

The middle of the currently on-going eclipse predicted by our 
Keplerian orbital solution 
is JD 2455136 $\pm$ 59; 2009 October 31.  The photometric 
prediction of mid-eclipse is JD 2455413; 2010 August 04 
(http://www.hposoft.com/Campaign09.html).  Therefore the spectroscopic mid-eclipse 
precedes the photometric mid-eclipse prediction by nine months.  Furthermore, 
the mid-eclipse times of previous eclipses predicted by the orbital solution
disagree with the mid-eclipse predictions established by photometry.

This would imply that the gravitating companion responsible for the
orbital motion is not the same as the extended structure responsible
for the eclipses, and that there is a positional offset between the
two, possibly due to a complex disk structure.  Alternatively,
perturbations in the radial velocities related to the short-period
oscillations could be affecting the orbital solution in subtle ways,
perhaps biasing the shape parameters ($e$ and $\omega$) on which the
predicted eclipse times depend rather strongly. We show below that the
amplitude of these oscillations is quite significant (roughly half of
the orbital amplitude). A combination of both effects is also
possible. If the spectroscopic orbit is biased, the observed times of
mid-eclipse as established from photometry could be used
simultaneously with the radial velocities to constrain the solution.
In the next sections we describe how we determine these times of mid
eclipse, and how we incorporate them into an alternate orbital
solution.

\section{MID-ECLIPSE TIMES}

Many estimates have been made of mid-eclipse times using a number of methods.
In general, these times are 
determined from the times of $1^{st}$, $2^{nd}$, $3^{th}$ and $4^{th}$
 contacts.  These times
have generally been established by extrapolating the ingress and egress light 
curves to the mean out-of-eclipse magnitude, often assuming that the eclipse
light curve is symmetrical.  Unfortunately, this
is not straightforward since there is considerable out-of-eclipse light 
variation due to the short period oscillation and, for some of the past eclipses,
 the contact points were not 
observationally accessible.  Furthermore, the ingress and egress light
curves are not symmetrical and not linear between $1^{st}$ and $2^{nd}$,
and $3^{th}$ and $4^{th}$
contacts.  Therefore, we have re-determined the mid-eclipse times
for all previous eclipses using the following procedure.  

We have collected photometry of $\epsilon$ Aurigae from
1842 to the present. In some cases the actual data were not published, so we obtained
them by digitizing the corresponding figures from the original publications.
All measurements are given in Table~\ref{tab:photometry} where columns 
give JD $- 2,\!400,\!000$, magnitude, flag for
linear fit (see below) and bibliographic code.  We note that the magnitude scale 
changed between the 1929 and
1956 eclipse from visually estimated magnitudes to $V$ magnitudes on the standard
Johnson system. The entire photometry table 
is available in the electronic version of the Journal for the benefit of future users.
  In Figure 2 we show a composite plot of the photometry.

For the 1875, 1902, 1929,
1956, and 1983 eclipses we use photometry from before and after the eclipses to
determine the mean out-of-eclipse magnitude,
over as
long a time period as possible to average out the short period
light variation of the star. We then fit a straight-line to the
``linear'' portion of the 
ingress and egress light curve.  The ``linear'' part of the light curves was
established by a careful examination of the photometric data for linearity
and, at times, excluding photometric data that was obviously inconsistent
with the bulk of the photometric observations, and giving preference to measurements
by observers showing the greatest internal consistency. The observations used in the 
linear fit are flagged with a ``1'' in column 3 of Table~\ref{tab:photometry}. The time 
differences established 
by the intersection 
of these ingress and egress linear fits 
and the mean out-of-eclipse magnitude establish a ``duration'' of each eclipse.
We note that the intersection points are not the same as the traditional initial and final
contacts nor is our ``duration'' the same as the traditional eclipse duration,
that is, the time from $1^{st}$ to $4^{th}$ contact.  But this procedure has the virtue that 
it can be determined more easily and objectively, and is much less susceptible 
to the short-period brightness oscillations in $\epsilon$ Aurigae than the traditional
method of estimating the contact times.
We then estimate the mid-eclipse time as the
average of the mid times between the fitted ingress and egress branches
at seven magnitude levels starting at the mean out-of-eclipse
magnitude and separated by $0.1$ magnitudes.  The results are shown in 
Table~\ref{tab:mid-eclipse} where
we give the slopes of the ingress and egress linear fits in magnitudes per day, the 
``duration,'' and our estimate of the mid-eclipse time and an error estimate.  

For the 1848 eclipse the egress was examined as described above. However, there is 
very little photometry before the eclipse and only
one observation during the ingress. 
We estimated the ``duration'' of the 1848 eclipse as the mean of the previous 
five eclipses, 668 days,  
and the slope of the ingress as the mean of the 1929, 1902, 1875 ingress slopes.  We then
use the procedure described above to estimate the mid-eclipse time.
   
We show in Figure 3 the eclipse photometry as well as the features described 
above for all previous eclipses.  The observations used for the linear fit are shown
as large open circles.

\section{COMBINED ORBITAL SOLUTION}

The eclipse timings constrain the ephemeris for $\epsilon$~Aurigae
(period and epoch) much more precisely than do the radial velocities.
In order to make use of this information, for the combined orbital
solution we have incorporated the eclipse times along with the radial
velocities into the least-squares fit, with their corresponding
observational errors.  We did this by predicting the times of eclipse
at each iteration based on the current spectroscopic elements, and
adding the residuals to those from the velocities in a $\chi^2$
sense. The timing errors for the six historical eclipses were
conservatively increased over their formal values by adding a fixed
amount in quadrature so as to achieve a reduced $\chi^2$ near unity
for these measurements. The proper amount (2.65 days) was found by
iterations, and the combined uncertainties reported in Table~4 include
this adjustment, and are believed to be realistic.

The orbital elements of our combined orbital solution are given in the last 
column of Table~\ref{tab:elements}. In Figure 4 we plot the individual observed
velocities as a function of orbital phase together with the velocity
curve calculated from our combined orbital solution.  The center-of-mass
velocity, $\gamma$, is shown as a horizontal dashed line. In Figure 1 we 
plot our orbital solution over all the radial velocities as a function of time.
According to our combined spectroscopic and photometric orbital solution, 
the predicted mid time of the eclipse currently underway, is 
JD $2455413.8 \pm 4.8$, corresponding to 2010 August 05.

\section{SHORT-TERM OSCILLATIONS}

It is clear from Figure 1 that there are short-term variations in the radial velocity
of $\epsilon$ Aurigae both inside and outside of the eclipses.  In addition, during 
the eclipse phases, considerable structure is observed in the spectral lines and 
multiples of several lines are observed. Presumably these features are due to 
structures in the occulting disk.  Various authors have speculated 
about the origin, nature and regularities of these oscillations, as well as 
the photometric variations. \citet{AF85}
found no regular oscillation from an examination of the Yerkes \citep{FSE29} and 
Mt. Wilson \citep{SPZ58} velocities and concluded that $\epsilon$ Aurigae is a non-radial
pulsator.       

Using CfA velocities outside the eclipse phase, we examined the
data to see if they can shed any light on the nature of the short-term
oscillations. The CfA velocity residuals from the spectroscopic orbit vary 
in a range of 5 to 20 \kms \ in no apparent pattern and the power spectrum of 
the residuals shows no clear periodicities.  The CfA velocity residuals, however, often
show well defined oscillations that last for only one or two cycles and are reminiscent 
of radial pulsation (Figure 5). 
These oscillations differ in
period from 75 to 175 days and have peak-to-peak amplitudes of 10 to 20 \kms.    

An auto-correlation of 488 velocities taken in the interval
JD 2447848--54579 shows a clear positive feature at about 
600 days (Figure 6).  This means that there is a correlation (about 0.5) between 
the modulation at a time $t$ and a time $t+600$.  Assuming the radial velocity 
residuals are coming from the motion of the stellar envelope facing us, 
this means that if the envelope is moving toward us then after
600 days it is likely, with a correlation of about 0.5, to move again toward 
us. There are less significant positive features in the auto-correlation 
at 300 and 450 days.

\acknowledgements 

We thank Perry Berlind, Joe Caruso, Michael Calkins and Gil Esquerdo 
for obtaining many of the spectroscopic observations used here.  
This research has made use of the SIMBAD database, operated at CDS, 
Strasbourg, France; and NASA's Astrophysics Data System Abstract Service.
We acknowledge with thanks the variable star observations from the AAVSO 
International Database contributed by observers worldwide and used in 
this research. We also thank Jeff Hopkins for his compilation of
photometric observations on his web site:
http://www.hposoft.com/Astro/PEP/EpsilonAurigae.html and his many years 
of photometric monitoring of $\epsilon$ Aurigae.

\begin{deluxetable}{cccl}
\tabletypesize{\scriptsize}
\def\th{\thinspace}
\def\kms{\ifmmode{\rm km\th s^{-1}}\else km\th s$^{-1}$\fi}
\tablewidth{0pt}
\tablecaption{Heliocentric Radial Velocities of $\epsilon$ Aurigae\label{tab:velocities}}
\tablehead{
\colhead{Julian Date}                   &
\colhead{Radial Velocity}               &
\colhead{error}                         &
\colhead{Reference}                     \\
\colhead{(HJD $-2,\!400,\!000)$}             &
\colhead{(\kms)}                        &
\colhead{(\kms)}                        &
\colhead{Code\tablenotemark{a}}         }
\startdata
    13932.     &      9.0  &    -    &   1, Lick  \\
    14987.     &      4.0  &    -    &   1, Adams \\
    14995.     &      4.0  &    -    &   1, Adams \\
    14996.     &      4.0  &    -    &   1, Adams \\
    15698.     &      5.3  &    -    &   1        \\
    16072.     &  $-$14.1  &    -    &   1        \\
\enddata
\tablenotetext{a}
{Reference:
(1) \citet{L24}; 
(2) \citet{CM28};
(3) \citet{A70};
(4) \citet{FSE29};
(5) \citet{SPZ58};
(6) \citet{AF85};
(7) \citet{BE86};
(8) \citet{P83};
(9) \citet{BLM86};
(10) \citet{C91};
(11) \citet{LS86};
(12) CfA}

\end{deluxetable}

\clearpage

\begin{deluxetable}{lcccccc}
\rotate
\tabletypesize{\scriptsize}
\def\th{\thinspace}
\def\kms{\ifmmode{\rm km\th s^{-1}}\else km\th s$^{-1}$\fi}
\tablewidth{0pt}
\tablecaption{Spectroscopic Orbital Elements of $\epsilon$ Aurigae\label{tab:elements}}
\tablehead{
\colhead{Element}                    &
\colhead{\citet{L24}}                &
\colhead{\citet{KSS37}}              &
\colhead{\citet{M62}}                &
\colhead{\citet{W70}}                &
\colhead{Keplerian Fit}              &
\colhead{Combined Fit\tablenotemark{a}}    }
\startdata
$P$ (days)                        & $9890$  & $9890$ (assumed) & $9890$ (assumed)        & $9890$ (assumed)    & $9882    \pm 17$    & $9896.0   \pm 1.6$   \\ 
$e$                               & $0.35$  & $0.33$   	       & $0.172  \pm 0.033$      & $0.200  \pm 0.034$  & $0.290   \pm 0.016$ & $0.227    \pm 0.011$ \\
$\omega$ (deg)                    & $319.7$ & $350$            & $347.8  \pm 15.8$  	  & $346.4  \pm 11.0$  & $29.8    \pm 3.1$   & $39.2     \pm 3.4$   \\
$T$ (HJD $-2,\!400,\!000$)        & $22512$ & $23827$  	       & $23441  \pm 402$        & $33346  \pm 278$    & $34425   \pm 76$    & $34723    \pm 80$    \\
$\gamma$ (\kms)                   & $-1.8$  & $-2.5$   	       & $-1.29  \pm 0.39$       & $-1.37  \pm 0.39$   & $-2.41   \pm 0.15$  & $-2.26    \pm 0.15$  \\
$K$ (\kms)                        & $14.8$  & $15.7$   	       & $14.71  \pm 0.53$       & $15.00  \pm 0.58$   & $14.43   \pm 0.27$  & $13.84    \pm 0.23$  \\
a$_1$$\sin{\it{i}}$ ($10^{6}$ km) & $1887$  & $2014$           & $1970$                  & $2000$              & $1876    \pm 30$    & $1835     \pm 29$    \\   
\it{f}(\it{m}) (\rm{M}$_{\sun}$)  & $2.7$   & $3.34$           & $3.12$                  & \nodata             & $2.69    \pm 0.13$  & $2.51     \pm 0.12$  \\
$\sigma$ (\kms)                   & \nodata & \nodata          & \nodata                 & \nodata             & $4.59$              & $4.63$               \\
$N$                               & $197$   & \nodata          & \nodata                 & \nodata             & $1014$              & $1020$               \\
Time Span (days)                  & $8583$  & $\sim 15340$     & \nodata                 & \nodata             & $40947$             & $40947$              \\
Cycles                            & $0.87$  & $\sim 1.55$      & \nodata                 & \nodata             & $4.1$               & $4.1$                \\
Major Source of Velocities        & Potsdam & Potsdam + Yerkes & Same + DAO + Mt Wilson  & Same                & Same + CfA          & Same +CfA            \\
\enddata
\tablenotetext{a}{For the Combined Fit we have combined velocity data with the times of
mid-eclipses established from photometry, Table~\ref{tab:mid-eclipse}.}

\end{deluxetable}

\clearpage

\begin{deluxetable}{cccc}
\tabletypesize{\scriptsize}
\def\th{\thinspace}
\def\kms{\ifmmode{\rm km\th s^{-1}}\else km\th s$^{-1}$\fi}
\tablewidth{0pt}
\tablecaption{Photometry of $\epsilon$ Aurigae\label{tab:photometry}}
\tablehead{
\colhead{Julian Date}                   &
\colhead{Magnitude}                     &
\colhead{Linear Fit}                    &
\colhead{Reference}                     \\
\colhead{(HJD $-2,400,000$)}            &
\colhead{}                              &
\colhead{Flag}                          &
\colhead{Code\tablenotemark{a}}         }
\startdata
    $-$6050.     &    3.30  &    0    &   2   \\
    $-$5538.     &    3.39  &    0    &   2   \\
    $-$5489.     &    3.15  &    0    &   1   \\
    $-$5430.     &    3.15  &    0    &   1   \\
    $-$5402.     &    3.31  &    0    &   2   \\
    $-$5338.     &    3.31  &    0    &   1   \\
\enddata
\tablenotetext{a}
{Observer, Reference:
(1) Heis, \citet{L1903}; 
(2) Argelander, \citet{L1903};
(3) Schmidt, \citet{L1912};
(4) Oudemans, \citet{L1903};
(5) Schonfeld, \citet{L1903};
(6) Schwab, \citet{L1903};
(7) Plassmann, \citet{L1903};
(8) Sawyer, \citet{L1903};
(9) Porro, \citet{L1903};
(10) Luizet, \citet{L1903};
(11) Prittwitz, \citet{L1903};
(12) Kopff, \citet{L1903};
(13) Gotz, \citet{L1903};
(14) Nijland, \citet{G1936};     
(15) Plassmann, \citet{G1936};
(16) Enebo, \citet{G1936};
(17) Wendell, \citet{G1936};
(18) Schiller, \citet{G1936};
(19) Lohnert, \citet{G1936};
(20) Scharbe, \citet{G1936};
(21) Mundler, \citet{G1936};
(22) Lau, \citet{G1936};
(23) Hornig, \citet{G1936};
(24) Menze, \citet{G1936};
(25) Guthnick, \citet{G1936};
(26) Johansson, \citet{G1936};
(27) Guthnick \& Pavel, \citet{G1936};
(28) Gadomski, \citet{G1936};
(29) Graff, \citet{G1936};
(30) Kordylewski, \citet{G1936};
(31) Gussow, \citet{G1936};
(32) Kukarkin, \citet{G1936};
(33) Beyer, \citet{G1936};
(34) Danjon, \citet{G1936};
(35) Jacchia, \citet{G1936};
(36) Pagaczewski, \citet{G1936};
(37) Stebbins \& Huffer, \citet{G1936};
(38) Tschernov, \citet{G1936};
(39) Mrazek, \citet{G1936};
(40) Dziewulski, \citet{G1936};
(41) Kopal, \citet{G1936};
(42) \citet{F1960};    
(43) \citet{L1959};
(44) \citet{G1970} plot digitized;
(45) \citet{T1957};
(46) \citet{A1960};
(47) \citet{HK1958};
(48) \citet{PF1986};
(49) \citet{JAPOA1983};
(50) \citet{CZ1987};
(51) \citet{FWZ1985};
(52) \citet{SN1987};
(53) \citet{TS2001};
(54) \citet{BAC1984};
(55) \citet{AF85};
(56) Hopkins, \citet{H2009};
(57) Dumont, \citet{H2009}; 
(58) Ingvarsson, \citet{H2009};
(59) AAVSO;
(60) \citet{W1959}}

\end{deluxetable}

\clearpage

\begin{deluxetable}{ccccccc}
\tabletypesize{\scriptsize}
\def\th{\thinspace}
\def\kms{\ifmmode{\rm km\th s^{-1}}\else km\th s$^{-1}$\fi}
\tablewidth{0pt}
\tablecaption{Mid-eclipse Parameters for $\epsilon$ Aurigae\label{tab:mid-eclipse}}
\tablehead{
\colhead{Eclipse}                       &
\colhead{Slope IN}                      &
\colhead{Slope OUT}                     &
\colhead{Duration}                      &
\colhead{Mean magnitude}                &
\colhead{Mid-eclipse}                   &
\colhead{Std. dev.}                     \\
\colhead{}                              &
\colhead{$dV$ per day}                  &
\colhead{$dV$ per day}                  &
\colhead{days}                          &
\colhead{out-of-eclipse}                &
\colhead{HJD}                           &
\colhead{days}                         }
\startdata
    1848.0  & $+$0.00363  &  $-$0.00566   &  668     & 3.277 & 2396041.  &   11     \\
    1875.2  & $+$0.00332  &  $-$0.00606   &  649     & 3.352 & 2405955.  &   15     \\
    1902.2  & $+$0.00369  &  $-$0.00355   &  700     & 3.399 & 2415827.7 &   2.9    \\
    1929.3  & $+$0.00387  &  $-$0.00366   &  695     & 3.329 & 2425726.8 &   3.1    \\
    1956.4  & $+$0.00593  &  $-$0.00583   &  645     & 3.003 & 2435624.6 &   2.7    \\
    1983.5  & $+$0.00569  &  $-$0.00927   &  651     & 3.008 & 2445507.1 &   7.8    \\
    2009-11 & $+$0.00670  &   \nodata     &  \nodata & 3.025 & \nodata   & \nodata  \\
\enddata
\end{deluxetable}

\clearpage

\newpage

\begin{figure}
\epsscale{0.8}
\plotone{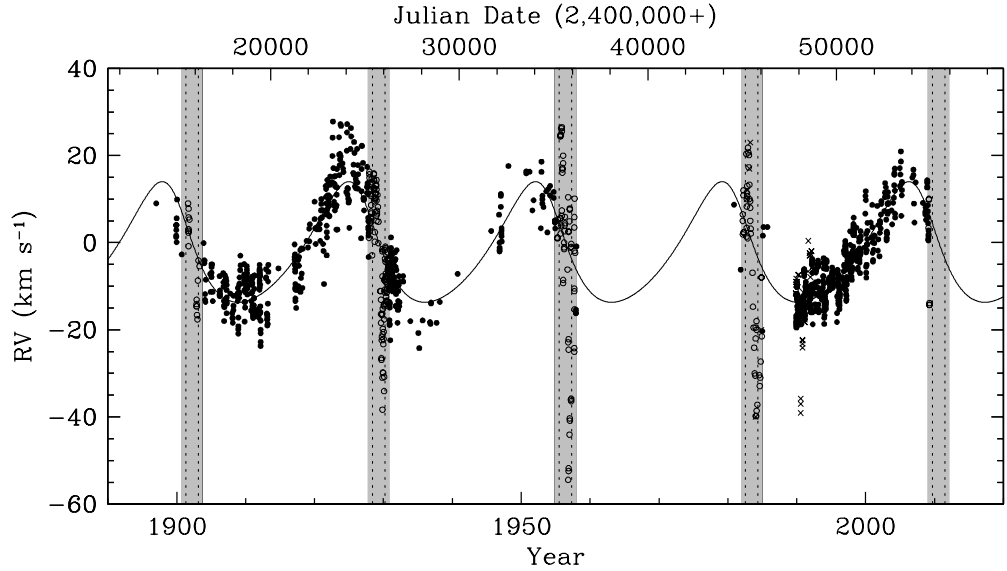}
\caption{History of the Radial Velocities of $\epsilon$ Aurigae.  Dashed lines indicate
eclipse ``duration''. The shaded area shows the time interval, ``duration''$\pm$ 200 days, with
the velocities excluded from the orbital solution shown as open circles. Filled circles are 
velocities used in the orbital solution and crosses are H$\alpha$ velocities.}
\label{fig:history}
\end{figure}

\begin{figure}
\epsscale{0.8}
\plotone{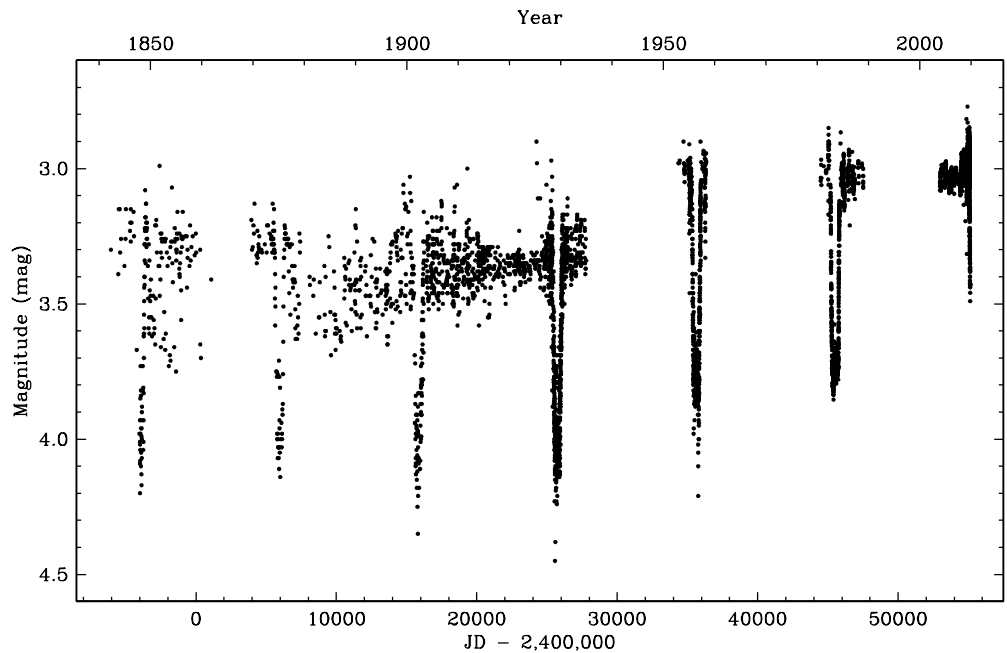}
\caption{History of the Photometry of $\epsilon$ Aurigae.  Note the change in the
magnitude scale between the early visual magnitudes and the more recent $V$ magnitudes.}
\label{fig:photometry}
\end{figure}

\begin{figure}
\epsscale{0.8}
\plotone{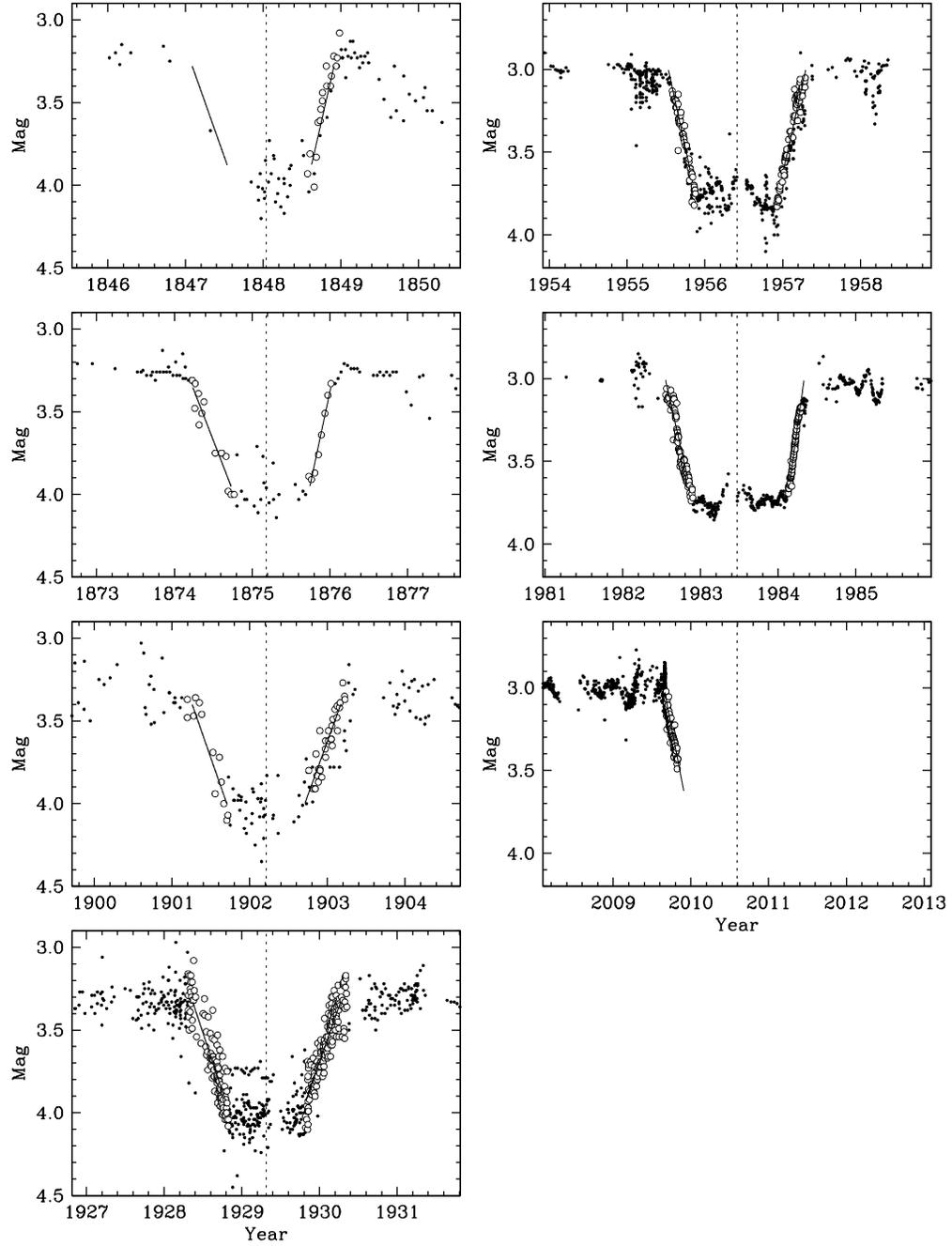}
\caption{Eclipse Light Curves of $\epsilon$ Aurigae.  Open circles
are observations used for linear fits, shown as solid lines, to the ingress 
and egress light variation during the eclipses.}
\label{fig:eclipses}
\end{figure}

\begin{figure}
\epsscale{0.8}
\plotone{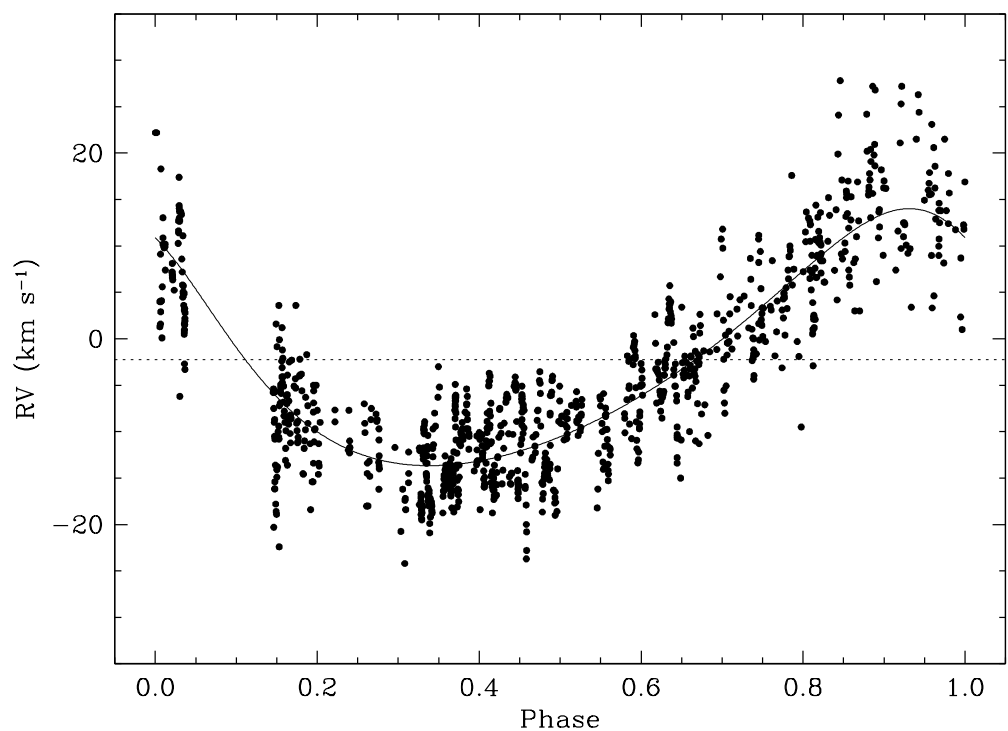}
\caption{Spectroscopic Orbital Solution for $\epsilon$ Aurigae.  
The center-of-mass velocity is indicated by the dashed line. Phase 0.0
corresponds to the time of periastron passage.}
\label{fig:orbit}
\end{figure}

\begin{figure}
\epsscale{0.8}
\plotone{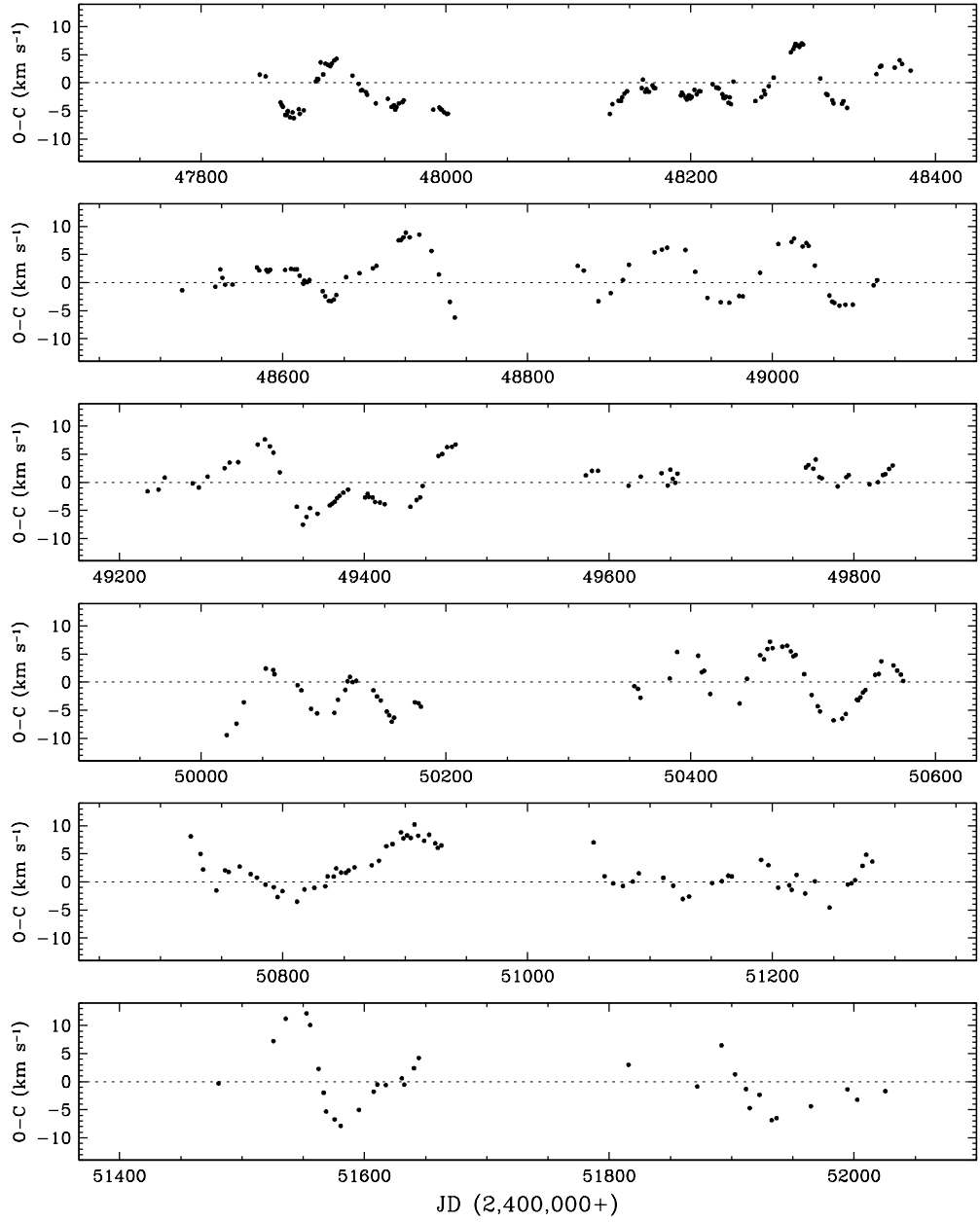}
\caption{Examples of Oscillations in the CfA Velocity Residuals.}
\label{fig:oscillations}
\end{figure}

\begin{figure}
\epsscale{0.8}
\plotone{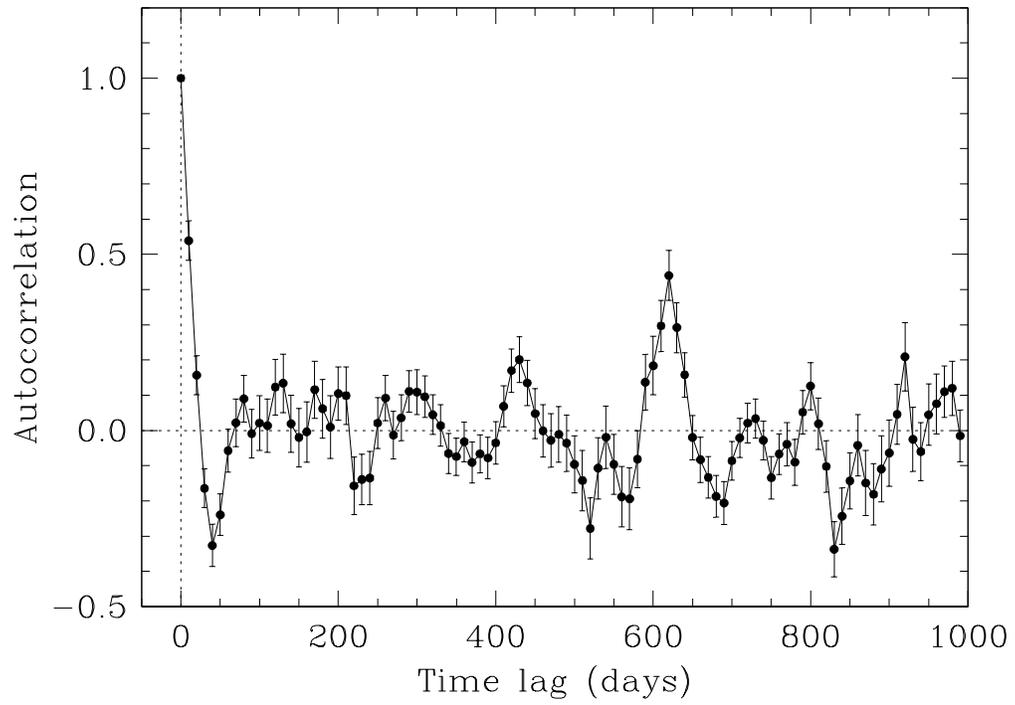}
\caption{Auto-correrlation of CfA Velocity Residuals.}
\label{fig:autocorrerlation}
\end{figure}

\end{document}